# Conversion Coefficients from Kerma to Ambient Dose and Personal Dose for X-Ray Spectra

**Thomas Otto**[a]

[a] *Technology Department, CERN*
*1211 Genève 23, Switzerland*
*E-mail*: thomas.otto@cern.ch

ABSTRACT: In radiation protection, the protection quantity for whole-body exposure is effective dose $E$. Effective dose cannot be measured and operational quantities have been introduced for dose measurement and calibration of dosimeters and survey instruments. To overcome some shortcomings of the presently used operational quantities, ICRU Report Committee 26 introduces two new quantities for whole body exposure, ambient dose $H^*$ for prospective dose assessment and personal dose $H_p(\alpha)$ for retrospective measurements with personal dosimeters. Dosimeters and survey instruments are calibrated in reference fields, realised with radioisotopes and with well-defined x-ray spectra. In this paper, the spectrum-averaged conversion coefficients from kerma in air $K_a$ to the new quantities ambient dose and personal dose are calculated and compared with the published coefficients for the present operational quantities. Especially at low energies ($E_{ph} < 40$ keV), the new quantities are significantly lower than the present ones, thus correcting a strong overestimate of effective dose.

# Contents

## 1. Introduction

In radiation protection, *protection quantities* are defined to express in a convenient way the radiation risk to persons. The protection quantity for exposure of the whole body is *effective dose*, $E$, defined as the sum of absorbed dose over 15 tissues (or organs) $D_{T,R}$, weighted for the relative sensitivity of the tissue ($w_T$) and the effectiveness of the radiation type ($w_R$) [1]:

$$E = \sum_R \sum_T w_T w_R D_{T,R} \qquad (1)$$

*Effective dose* is defined over the extended volume of the whole body and not directly measurable. Numerical values of $E$ and conversion coefficients from field quantities (such as fluence $\phi$ or kerma $K_a$) are evaluated with Monte-Carlo radiation transport calculations on mathematical anthropomorphic phantoms [2],[3].

The International Commission on Radiation Units and Measurements (ICRU) defines measurable *operational quantities* for radiation protection. Presently, two operational quantities are defined for dose assessment of whole-body exposures: *ambient dose equivalent* $H^*(10)$ for prospective assessments of radiation exposure at workplaces and in the environment with survey instruments and *personal dose equivalent* $H_p(10,\alpha)$ for retrospective assessment of dose received by a person with personal dosimeters. Both quantities have in common that they are evaluated in 10 mm depth within a reference phantom. Conversion coefficients to the operational quantities are evaluated by radiation transport calculation in the ICRU sphere or in slab- and cylinder phantoms, the latest numerical values have been published jointly in ICRP Report 74 [4] and ICRU Report 57 [5].

The operational quantities $H^*(10)$ and $H_p(10, \alpha)$ have a number of shortcomings rooted in their definition, because a measurement in a reference depth of 10 mm cannot properly represent the complexity of the human body with organs seated at different depths. Figure 1 shows the comparison between effective dose $E$ for photons in Anterior-Posterior (AP) orientation (irradiation from the front) and personal dose equivalent $H_p(10,0^O)$ evaluated with the kerma-approximation as in the official tabulations [4],[5] and with transport of secondary particles. The energy dependence of ambient dose equivalent $H^*(10)$ is very similar and would be indistinguishable on the scale of the plot. It is visible that the operational quantities for photons defined in 10 mm depth

- overestimate the protection quantity for low photon energies ($E_{ph} < 40$ keV)
- overestimate the protection quantity for high photon energies ($E_{ph} > 3$ MeV) in the kerma-approximation
- underestimate the protection quantity for high photon energies ($E_{ph} > 3$ MeV) when correctly calculated with secondary particle transport.

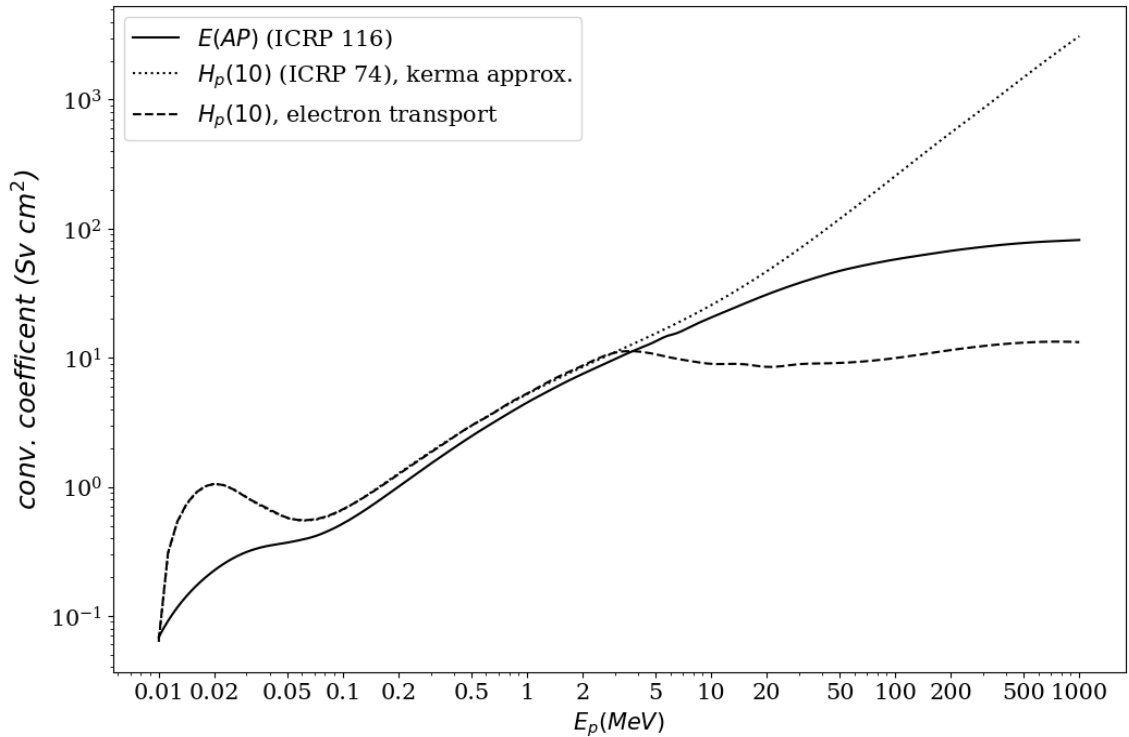

**Figure 1.** Effective dose per unit fluence in AP orientation $E$(AP) (continuous line), personal dose equivalent $H_p(10,0^0)$ per unit fluence as published, calculated in kerma-approximation (dotted line), personal dose equivalent $H_p(10,0^0)$ per unit fluence calculated with full electron transport (dashed line). On this scale, the corresponding curves for ambient dose equivalent $H^*(10)$ are indistinguishable from the ones for $H_p(10,0^0)$.

The ICRU has tasked Report Committee (RC) 26 to propose a redefinition of operational quantities [6], [7]. The committee proposes to introduce operational quantities directly based on the corresponding protection quantity. For the prospective assessment of whole-body doses at workplaces and in the environment, ambient dose $H^*$ is defined for a specified particle type and energy as the maximum value of effective dose $E$ for different orientations of the radiation field (AP, LLAT, RLAT, PA, ROT, ISO) with respect to the body. For retrospective assessments of whole-body dose received, and for the calibration of personal dosimeters, the quantity personal dose $H_p(\alpha)$ is defined for an incident angle $\alpha$ normal to the vertical axis of the body. The conversion coefficients for the operational quantities $H^*$ and $H_p(\alpha)$ are calculated within the same numerical phantoms as the protection quantities [2]. By definition, their numerical values are equal to those of the protection qualities for certain energies and angles. $H_p(0°)$ is numerically equivalent to $E$(AP). For photons, $H^*$ is numerically equivalent to $E$(AP) and to $H_p(0°)$ for energies $E_{ph} \leq 6$ MeV.

Figure 2a shows the energy dependence of the conversion coefficients from kerma in air $K_a$ to ambient dose equivalent $h_k^*(10,E)$ and to ambient dose $h_k^*(E)$ for the energy range of interest. For nearly all energies, $h_k^*(10,E) > h_k^*(E)$, with exception for very low energies (Figure 2b). The difference at very low energies is due to the contribution of the tissue 'skin' to effective dose $E$ and thus to $H^*$ and $H_p(\alpha)$: in contrast to a target at 10 mm depth, skin is receiving dose at even the lowest photon energies.

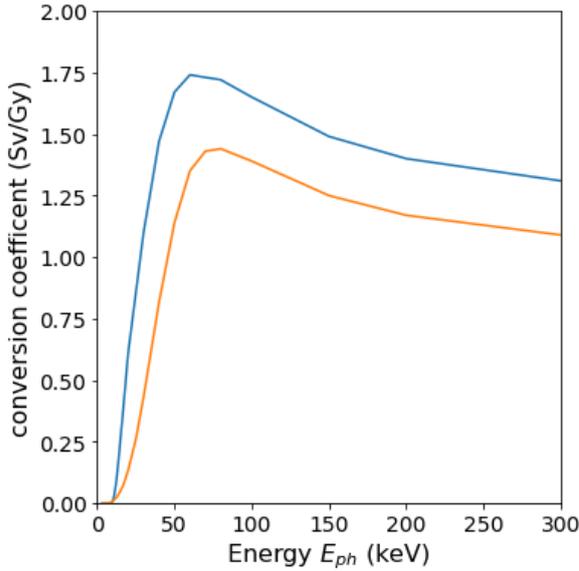 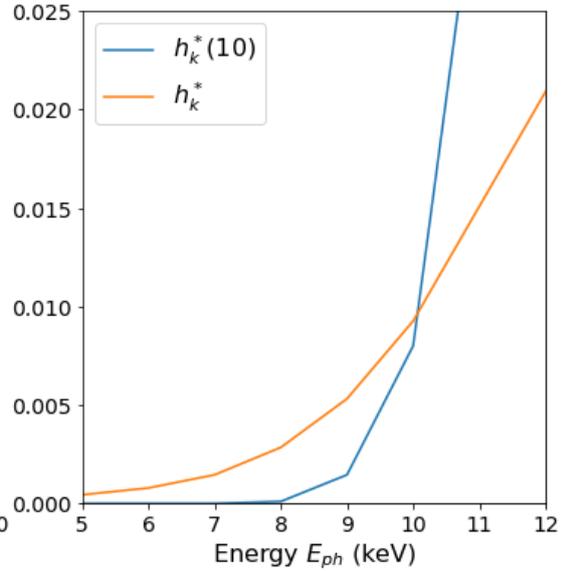

**Figure 2a.** Conversion coefficients from photon kerma in free air $K_a$ to ambient dose equivalent $H^*(10)$ (blue line) and ambient dose $H^*$ (orange line). $h_k^*(10,E) > h_k^*(E)$ for most of the energy range

**Figure 2b.** As figure 2a, in the energy range between 5 keV and 12 keV. $h_k^*(10,E) < h_k^*(E)$ for $E_{ph} < 10$ keV

From the relation between the monoenergetic conversion coefficients one can expect that the conversion coefficients for x-ray spectra are higher for the presently used quantities $H^*(10)$ and $H_p(10,\alpha)$ than for the proposed quantities $H^*$ and $H(\alpha)$, with exception for the qualities with the lowest tube potentials.

## 2. X-ray spectra for calibration of dosimeters, survey instruments and monitors

Radiation protection dosimeters and survey instruments for photons are calibrated with reference radiations from X-ray generators and from radioisotope sources, described in standard ISO 4037-1 [8]. Different series of x-ray spectra are defined by the x-ray tube potential, the filtration and the 1st and 2nd half-value layers of the spectrum. The narrow-spectrum series (N-series), is frequently used for calibration of radiation protection instruments and dosimeters, but the standard also lists a wide-spectrum series (W-series), a high air-kerma series (H-series) and a low air-kerma series (L-series). A German standard, DIN 6818-1 [9], defines series of x-ray spectra with identical definition, but for a few more tube voltages (A-, B-, C-Series).

X-ray qualities for the calibration of monitoring instruments in radiation diagnostics are defined in an analogous way in the international standard IEC 61267 [10]. It contains the definitions for the RQR-, RQA and RQT-series as surrogates for diagnostic radiation fields. The spectra are shaped such that they approximate x-ray spectra emitted by diagnostic units. While the radiation qualities IEC 61267 are not used primarily for the calibration of radiation protection instruments, they are of interest to assess the exposure of medical personnel during diagnostic procedures and technical personnel during type testing and calibration.

Kerma-to-dose conversion coefficients for these spectra are tabulated for the present operational quantities $H_p(10,\alpha)$ and $H^*(10)$ in [11], [12]. Generally, the conversion coefficient $c_k(S)$ from kerma to a quantity $C$ for a radiation quality $S$ with fluence spectrum $\phi_{E_p}^S(E_p)$ can be calculated by the expression

$$c_k(S) = \frac{\int \phi_{E_p}^S(E_p) \, k_\phi(E_p) \, c_k(E_p) dE_p}{\int \phi_{E_p}^S(E_p) \, k_\phi(E_p) dE_p} , \qquad (2)$$

where $c_k(E_p)$ are the monoenergetic conversion coefficients from kerma to quantity $C$. The integral is taken over the energy range covered by the fluence spectrum. The German National Standards Laboratory PTB makes fluence spectra $\phi_{E_p}^S(E_p)$ and kerma-weighted fluence spectra $\phi_{E_p}^S(E_p) \cdot k_\phi(E_p)$ for most of the radiation qualities listed in [12] available on its website [13]. With these data, the conversion coefficients from kerma in air $K_a$ to $H^*$, $H_p$ were calculated for the series of x-ray qualities for calibration of radiation protection instruments and diagnostic monitors according to Equation 2.

## 3. Calculation of conversion coefficients

For the numerical evaluation of (2) with tabulated data the integrals must be discretised. The x-ray spectra from PTB [13] are tabulated with a resolution between 0.2 keV and 1.0 keV. The values of the conversion coefficients from kerma in air $K_a$ to $H^*$ and $H_p$ can be calculated from the conversion coefficients from fluence published in [6] and kerma-to-fluence conversion coefficients from [4],[5]. Tables with conversion coefficients from Kerma to $K_a$ to $H^*$ and $H_p$ will be published in the forthcoming ICRU report. The conversion coefficients were interpolated by a cubic spline function on a logarithmic energy scale to match the resolution of the x-ray spectra. For this purpose, the `scipy` [14] library routine `interpolate.interp1d` was used. The integrals were evaluated by Simpson's rule with the library routine `integrate.simps`.

As a quality check, the conversion coefficients for $H^*(10)$ and $H_p(10, \alpha)$ were calculated with the algorithm described above and compared with the results published in [12]. The differences are generally lower than 1 percent. Only for x-ray spectra with the lowest tube voltages, for example N10, deviations of up to 4.5 % are observed, owing to the different interpolation- and integration algorithms.

## 4. Results

In this section, the conversion coefficients from kerma in air $K_a$ to Ambient dose $h^*(S)$ and to personal dose $h_p(S, \alpha)$ are presented for each spectrum $S$ from a series of x-ray qualities (spectra). They are quoted with two significant digits because they are based on monoenergetic coefficients [6] with this precision. In the energy range of interest, the conversion coefficients for ambient dose $h^*(S)$ are numerically equivalent to those for personal dose $h_p(S, 0°)$ and are quoted in the same column with the title $h_p^*(S,0)$. Conversion coefficients with a numerical value of less than $10^{-4}$ are not listed.

## 4.1 Radiation Qualities from ISO 4037-1

The standard ISO 4037-1 contains spectra for the calibration of radiation dosimeters and survey instruments. The narrow-spectrum series of x-ray qualities is frequently used for type-testing and calibration of radiation protection dosimeters and survey instruments. For convenience, conversion coefficients for the radioisotope spectra S-Cs and S-Co from the standard for the radiation from $^{137}$Cs and $^{60}$Co were added to table 1.

**Table 1:** Spectrum-averaged conversion coefficients from kerma to the new operational quantities Ambient dose $h_k^*(S, 0°)$, and personal dose $h_p(S, 0°)$ for radiation spectra from the series of Narrow-spectrum (N) qualities from [8]. The kerma-weighted fluence spectra are taken from [13]. Conversion coefficients for radioisotope qualities S-Cs and S-Co are added.

| Spectrum S | $E_{avg}$ (keV) | $h_p^*(S,0)$ (Sv/Gy) | $h_p(S,15)$ (Sv/Gy) | $h_p(S,30)$ (Sv/Gy) | $h_p(S,45)$ (Sv/Gy) | $h_p(S,60)$ (Sv/Gy) | $h_p(S,75)$ (Sv/Gy) | $h_p(S,90)$ (Sv/Gy) | $h_p(S,180)$ (Sv/Gy) | $h_p(S,\text{ROT})$ (Sv/Gy) | $h_p(S,\text{ISO})$ (Sv/Gy) |
|---|---|---|---|---|---|---|---|---|---|---|---|
| N-10 | 8.5 | 0.0039 | 0.0040 | 0.0038 | 0.0033 | 0.0028 | 0.0021 | 0.0013 | 0.0017 | 0.0023 | 0.0020 |
| N-15 | 12.3 | 0.022 | 0.022 | 0.020 | 0.017 | 0.013 | 0.0094 | 0.0057 | 0.0033 | 0.0099 | 0.0084 |
| N-20 | 16.3 | 0.061 | 0.060 | 0.056 | 0.048 | 0.038 | 0.026 | 0.016 | 0.0064 | 0.027 | 0.022 |
| N-25 | 20.3 | 0.13 | 0.13 | 0.12 | 0.10 | 0.083 | 0.057 | 0.035 | 0.017 | 0.058 | 0.047 |
| N-30 | 24.6 | 0.23 | 0.23 | 0.22 | 0.19 | 0.15 | 0.11 | 0.07 | 0.05 | 0.11 | 0.089 |
| N-40 | 33 | 0.53 | 0.52 | 0.49 | 0.43 | 0.36 | 0.26 | 0.17 | 0.19 | 0.28 | 0.22 |
| N-60 | 48 | 1.03 | 1.00 | 0.96 | 0.85 | 0.73 | 0.55 | 0.40 | 0.56 | 0.62 | 0.49 |
| N-80 | 65 | 1.37 | 1.35 | 1.29 | 1.16 | 1.00 | 0.78 | 0.58 | 0.88 | 0.89 | 0.71 |
| N-100 | 83 | 1.42 | 1.41 | 1.34 | 1.22 | 1.06 | 0.85 | 0.64 | 0.98 | 0.96 | 0.77 |
| N-120 | 100 | 1.38 | 1.37 | 1.30 | 1.20 | 1.05 | 0.85 | 0.64 | 0.97 | 0.95 | 0.77 |
| N-150 | 118 | 1.33 | 1.32 | 1.24 | 1.16 | 1.02 | 0.84 | 0.63 | 0.94 | 0.93 | 0.75 |
| N-200 | 165 | 1.22 | 1.21 | 1.15 | 1.07 | 0.96 | 0.80 | 0.61 | 0.89 | 0.87 | 0.70 |
| N-250 | 207 | 1.16 | 1.15 | 1.10 | 1.03 | 0.93 | 0.78 | 0.60 | 0.87 | 0.84 | 0.69 |
| N-300 | 248 | 1.12 | 1.12 | 1.07 | 1.01 | 0.91 | 0.78 | 0.60 | 0.85 | 0.82 | 0.68 |
| S-Cs | 662 | 1.01 | 1.02 | 0.99 | 0.95 | 0.89 | 0.78 | 0.65 | 0.84 | 0.81 | 0.70 |
| S-Co | 1225 | 1.00 | 1.00 | 0.98 | 0.95 | 0.90 | 0.82 | 0.72 | 0.87 | 0.84 | 0.74 |

Figures 3- 6 in this section show for selected radiation qualities $S$ the fluence spectrum $\phi_{E_p}^S(E_p)$, the fluence spectrum weighted by the conversion coefficients to ambient dose equivalent $\phi_{E_p}^S(E_p) \cdot k_\phi(E_p) \cdot h_k^*(10, E)$ and to ambient dose $\phi_{E_p}^S(E_p) \cdot k_\phi(E_p) \cdot h_k^*(E)$. The fluence spectrum is normalised to unit fluence (1 cm$^{-2}$) and both dose-weighted spectra are normalised to the integral of the present operational quantity $\int \phi_{E_p}^S(E_p) k_\phi(E_p) h_k^*(10, E) dE_p$. The purpose of these figures is to illustrate how different parts of the spectrum contribute to the operational quantity, and how the change to the new quantity will lead to lower conversion coefficients, especially for low x-ray tube voltages.

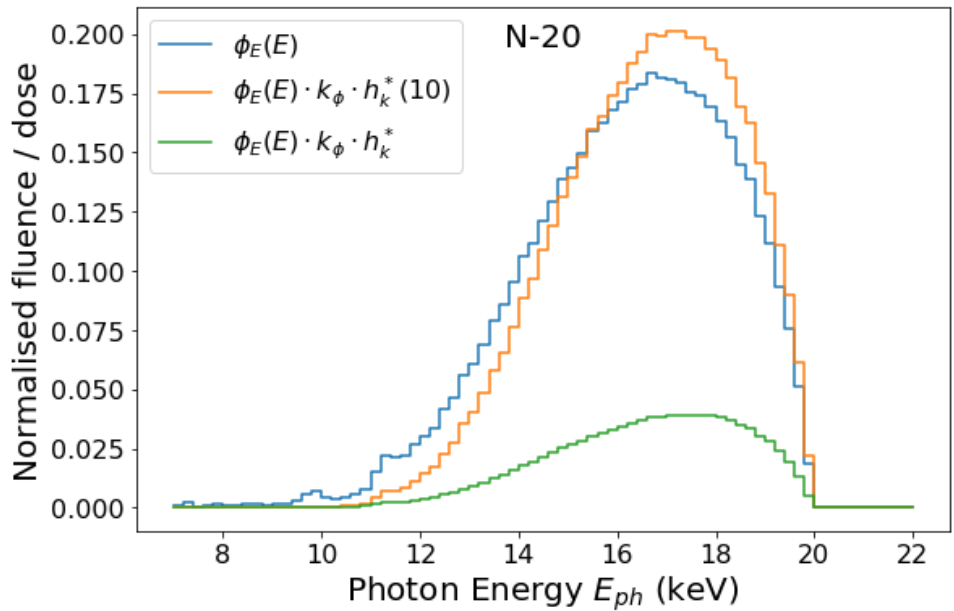

**Figure 3:** Fluence spectrum (blue), ambient dose-equivalent weighted fluence spectrum (orange) and ambient dose weighted fluence spectrum (green) for the N-20 quality from the narrow (N) series. The spectrum weighted with the new conversion coefficients has a strongly reduced amplitude with respect to the present dose-equivalent weighed spectrum. This accounts for the significant overestimation of effective dose by the present operational quantities.

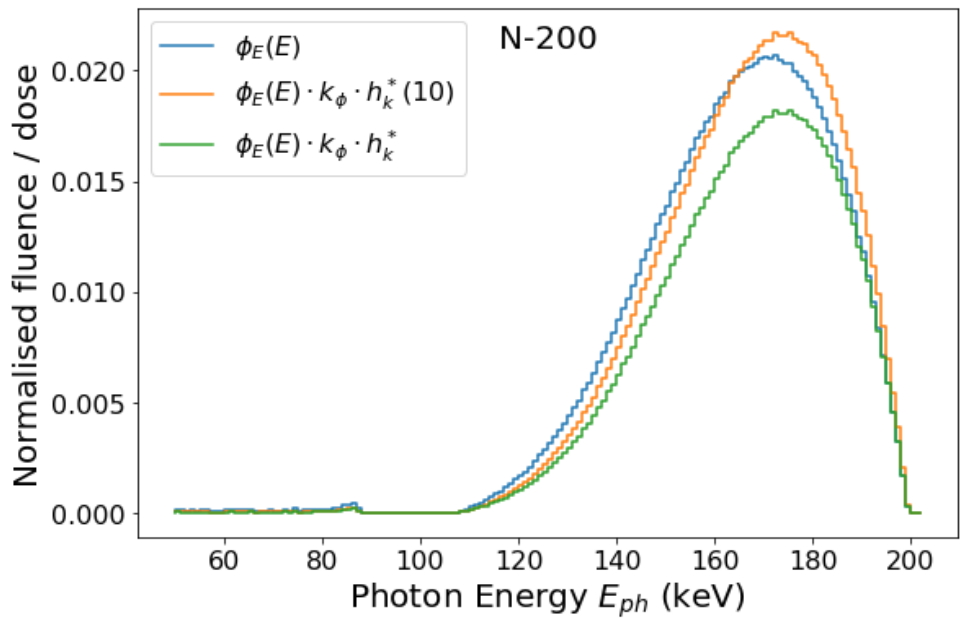

**Figure 4:** Fluence spectrum (blue), ambient dose-equivalent weighted fluence spectrum (orange) and ambient dose weighted fluence spectrum (green) for the N-200 quality from the narrow (N) series. In this energy range, the overestimation of effective dose by the present operational quantity is about 20 %, reflected in the smaller difference between the dose-weighted spectra.

**Table 2:** Spectrum-averaged conversion coefficients from kerma to the new operational quantities Ambient dose $h_k^*(S, 0°)$, and personal dose $h_p(S, 0°)$ for radiation spectra from the series of Low air-kerma rate (L) qualities from [8]. The kerma-weighted fluence spectra are taken from [13].

| Spectrum $S$ | $E_{\text{avg}}$ (keV) | $h_k^*(S,0)$ (Sv/Gy) | $h_p(S,15)$ (Sv/Gy) | $h_p(S,30)$ (Sv/Gy) | $h_p(S,45)$ (Sv/Gy) | $h_p(S,60)$ (Sv/Gy) | $h_p(S,75)$ (Sv/Gy) | $h_p(S,90)$ (Sv/Gy) | $h_p(S,180)$ (Sv/Gy) | $h_p(S,\text{ROT})$ (Sv/Gy) | $h_p(S,\text{ISO})$ (Sv/Gy) |
|---|---|---|---|---|---|---|---|---|---|---|---|
| L-10 | 9.0 | 0.0053 | 0.0053 | 0.0050 | 0.0043 | 0.0036 | 0.0026 | 0.0016 | 0.0020 | 0.0029 | 0.0025 |
| L-20 | 17.3 | 0.079 | 0.077 | 0.072 | 0.062 | 0.049 | 0.034 | 0.021 | 0.008 | 0.034 | 0.028 |
| L-30 | 26.7 | 0.30 | 0.30 | 0.28 | 0.25 | 0.20 | 0.14 | 0.091 | 0.073 | 0.15 | 0.12 |
| L-35 | 30.4 | 0.43 | 0.42 | 0.40 | 0.35 | 0.29 | 0.21 | 0.14 | 0.14 | 0.22 | 0.18 |
| L-55 | 47.8 | 1.05 | 1.02 | 0.98 | 0.88 | 0.75 | 0.57 | 0.41 | 0.57 | 0.64 | 0.51 |
| L-70 | 60.6 | 1.34 | 1.31 | 1.25 | 1.12 | 0.96 | 0.75 | 0.56 | 0.84 | 0.85 | 0.68 |
| L-100 | 86.8 | 1.42 | 1.41 | 1.34 | 1.22 | 1.07 | 0.86 | 0.64 | 0.98 | 0.97 | 0.77 |
| L-125 | 109.4 | 1.36 | 1.35 | 1.27 | 1.19 | 1.04 | 0.85 | 0.64 | 0.96 | 0.94 | 0.76 |
| L-170 | 148.5 | 1.25 | 1.24 | 1.18 | 1.09 | 0.98 | 0.81 | 0.62 | 0.91 | 0.88 | 0.72 |
| L-210 | 184.6 | 1.19 | 1.18 | 1.13 | 1.05 | 0.94 | 0.79 | 0.61 | 0.88 | 0.85 | 0.69 |
| L-240 | 211.4 | 1.16 | 1.15 | 1.10 | 1.03 | 0.93 | 0.78 | 0.60 | 0.86 | 0.84 | 0.68 |

**Table 3:** Spectrum-averaged conversion coefficients from kerma to the new operational quantities Ambient dose $h_k^*(S)$, and personal dose $h_p(S, 0°)$ for radiation spectra from the series of Wide-spectrum (W) qualities from [8]. The kerma-weighted fluence spectra are taken from [13].

| Spectrum $S$ | $E_{\text{avg}}$ (keV) | $h_p(S,0)$ (Sv/Gy) | $h_p(S,15)$ (Sv/Gy) | $h_p(S,30)$ (Sv/Gy) | $h_p(S,45)$ (Sv/Gy) | $h_p(S,60)$ (Sv/Gy) | $h_p(S,75)$ (Sv/Gy) | $h_p(S,90)$ (Sv/Gy) | $h_p(S,180)$ (Sv/Gy) | $h_p(S,\text{ROT})$ (Sv/Gy) | $h_p(S,\text{ISO})$ (Sv/Gy) |
|---|---|---|---|---|---|---|---|---|---|---|---|
| W-30* | 22.9 | 0.18 | 0.17 | 0.15 | 0.12 | 0.083 | 0.052 | 0.033 | 0.085 | 0.068 | 0.18 |
| W-40* | 29.8 | 0.37 | 0.35 | 0.31 | 0.25 | 0.18 | 0.12 | 0.11 | 0.19 | 0.15 | 0.37 |
| W-60 | 44.8 | 0.87 | 0.84 | 0.75 | 0.63 | 0.48 | 0.34 | 0.46 | 0.53 | 0.42 | 0.90 |
| W-80 | 56.5 | 1.16 | 1.11 | 1.00 | 0.85 | 0.66 | 0.48 | 0.71 | 0.75 | 0.59 | 1.19 |
| W-110 | 79.1 | 1.38 | 1.32 | 1.20 | 1.04 | 0.83 | 0.62 | 0.95 | 0.94 | 0.75 | 1.40 |
| W-150 | 104.2 | 1.34 | 1.27 | 1.17 | 1.03 | 0.84 | 0.63 | 0.95 | 0.93 | 0.75 | 1.35 |
| W-200 | 137.5 | 1.26 | 1.20 | 1.11 | 0.99 | 0.82 | 0.62 | 0.92 | 0.90 | 0.72 | 1.27 |
| W-250 | 172.3 | 1.20 | 1.14 | 1.06 | 0.95 | 0.80 | 0.61 | 0.89 | 0.86 | 0.70 | 1.21 |
| W-300 | 205.4 | 1.15 | 1.11 | 1.03 | 0.93 | 0.79 | 0.61 | 0.87 | 0.84 | 0.69 | 1.16 |

* These qualities are defined as *B30* and *B40* in [9].

**Table 4:** Spectrum-averaged conversion coefficients from kerma to the new operational quantities Ambient dose $h_k^*(S, 0°)$, and personal dose $h_p(S, 0°)$ for radiation spectra from the series of High air-kerma rate (H) qualities from [8]. The kerma-weighted fluence spectra are taken from [13].

| Spectrum S | $E_{avg}$ (keV) | $h_p^*(S,0)$ (Sv/Gy) | $h_p(S,15)$ (Sv/Gy) | $h_p(S,30)$ (Sv/Gy) | $h_p(S,45)$ (Sv/Gy) | $h_p(S,60)$ (Sv/Gy) | $h_p(S,75)$ (Sv/Gy) | $h_p(S,90)$ (Sv/Gy) | $h_p(S,180)$ (Sv/Gy) | $h_p(S,\text{ROT})$ (Sv/Gy) | $h_p(S,\text{ISO})$ (Sv/Gy) |
|---|---|---|---|---|---|---|---|---|---|---|---|
| H-20 | 13.1 | 0.024 | 0.023 | 0.021 | 0.018 | 0.014 | 0.010 | 0.0062 | 0.0034 | 0.011 | 0.0089 |
| H-30 | 19.5 | 0.093 | 0.091 | 0.085 | 0.074 | 0.059 | 0.041 | 0.026 | 0.014 | 0.042 | 0.034 |
| H-40* | 25.4 | 0.21 | 0.21 | 0.19 | 0.17 | 0.14 | 0.10 | 0.06 | 0.05 | 0.10 | 0.082 |
| H-60 | 38.0 | 0.58 | 0.57 | 0.54 | 0.48 | 0.40 | 0.30 | 0.20 | 0.25 | 0.32 | 0.26 |
| H-80* | 48.8 | 0.90 | 0.88 | 0.84 | 0.75 | 0.64 | 0.49 | 0.35 | 0.48 | 0.54 | 0.43 |
| H-100 | 57.3 | 1.09 | 1.07 | 1.02 | 0.91 | 0.78 | 0.61 | 0.44 | 0.64 | 0.68 | 0.54 |
| H-150* | 78.0 | 1.32 | 1.30 | 1.24 | 1.13 | 0.98 | 0.78 | 0.58 | 0.87 | 0.88 | 0.70 |
| H-200 | 99.3 | 1.32 | 1.31 | 1.24 | 1.14 | 1.00 | 0.81 | 0.61 | 0.92 | 0.90 | 0.73 |
| H-250 | 1231.5 | 1.28 | 1.27 | 1.21 | 1.12 | 0.99 | 0.81 | 0.62 | 0.91 | 0.89 | 0.72 |
| H-280 | 144.6 | 1.24 | 1.23 | 1.18 | 1.09 | 0.97 | 0.81 | 0.61 | 0.90 | 0.88 | 0.71 |
| H-300 | 143.2 | 1.24 | 1.23 | 1.17 | 1.09 | 0.97 | 0.81 | 0.61 | 0.90 | 0.88 | 0.71 |

\* These qualities are defined as *C30, C80* and *C150* in [9].

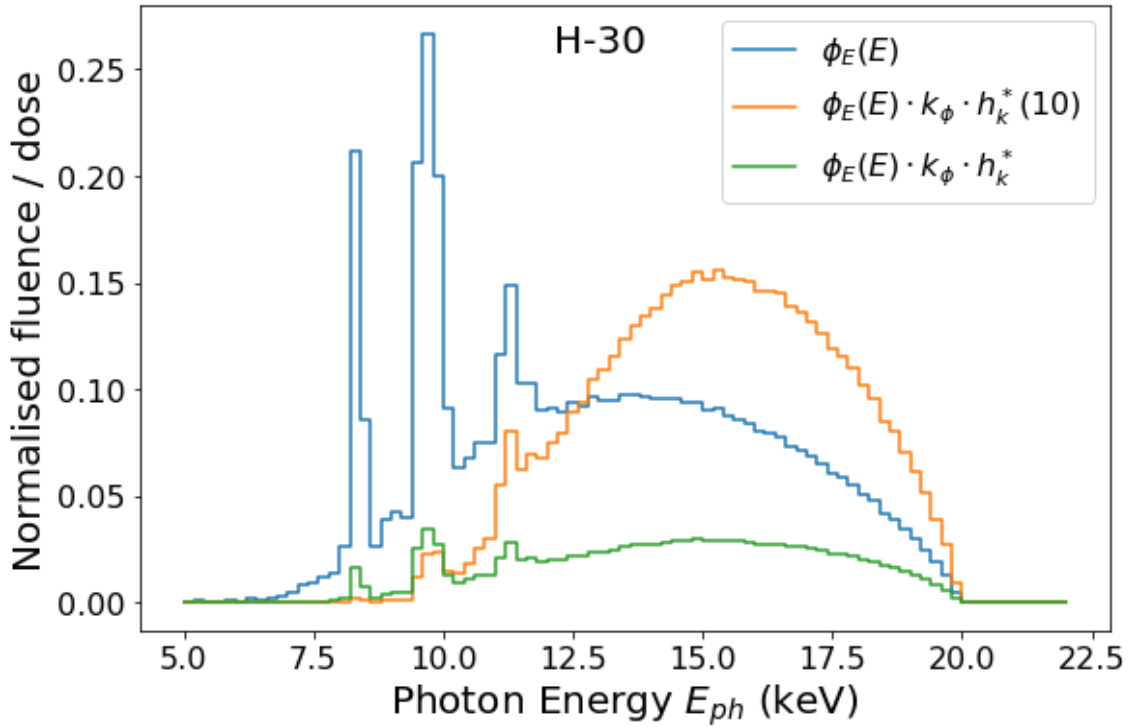

**Figure 4:** Fluence spectrum (blue), ambient dose-equivalent weighted fluence spectrum (orange) and ambient dose weighted fluence spectrum (green) for the H-30 quality from the high air-kerma (H) series. The spectrum weighted with the new dose conversion coefficients has a strongly reduced amplitude with respect to the present dose-equivalent weighted spectrum. This accounts for the significant overestimation of effective dose by the present operational quantities.
.

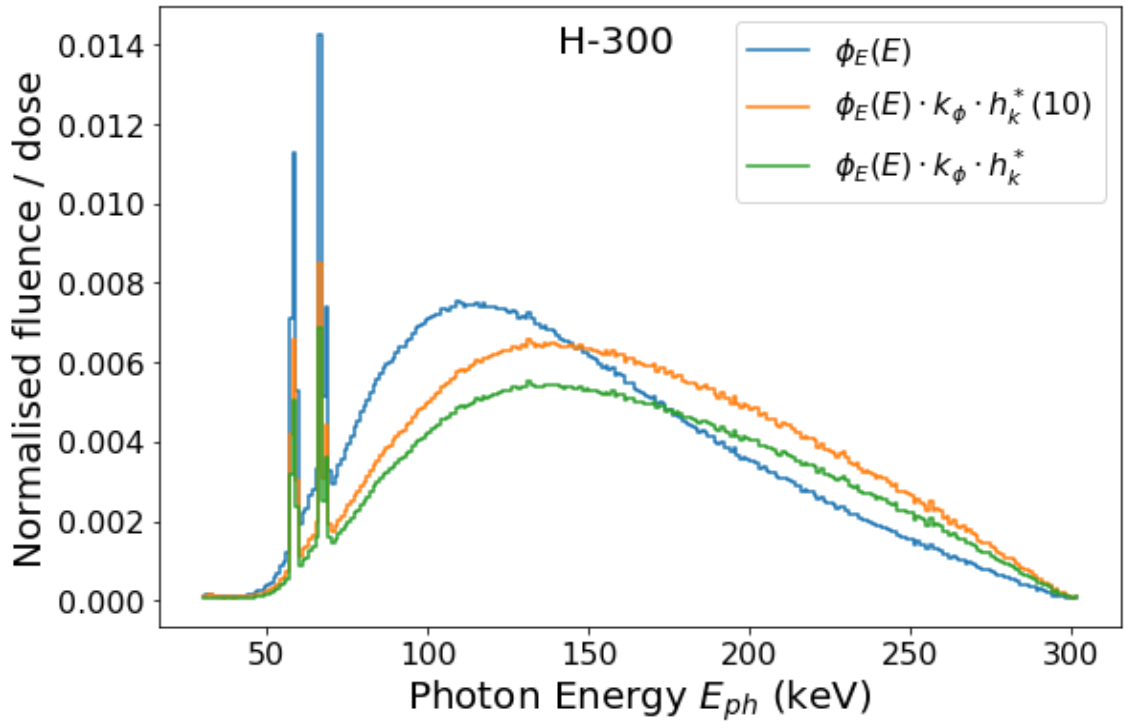

**Figure 6:** Fluence spectrum (blue), ambient dose-equivalent weighted fluence spectrum (orange) and ambient dose weighted fluence spectrum (green) for the H-300 quality from the high air-kerma (H) series. In this energy range, the overestimation of effective dose by the present operational quantity is about 20 %, reflected in the smaller difference between the dose-weighted spectra.

### 4.2 Radiation Qualities from IEC 61267

The standard IEC 61267 describes x-Ray qualities for the calibration of monitoring instruments for diagnostic radiation units. They are summarized in the RQR-, RQA- and RQT series. The catalogue of Ankerhold contains a number of these spectra which were used to calculate spectrum-averaged conversion coefficients [12].

**Table 5:** Spectrum-averaged conversion coefficients from kerma to the new operational quantities ambient dose $h_k^*(S, 0°)$, and personal dose $h_p(S, 0°)$ for radiation spectra from the series of diagnostic radiation qualities from [10]. The kerma-weighted fluence spectra are taken from [13].

| Spectrum $S$ | $E_{avg}$ (keV) | $h_p^*(S,0)$ (Sv/Gy) | $h_p(S,15)$ (Sv/Gy) | $h_p(S,30)$ (Sv/Gy) | $h_p(S,45)$ (Sv/Gy) | $h_p(S,60)$ (Sv/Gy) | $h_p(S,75)$ (Sv/Gy) | $h_p(S,90)$ (Sv/Gy) | $h_p(S,180)$ (Sv/Gy) | $h_p(S,\text{ROT})$ (Sv/Gy) | $h_p(S,\text{ISO})$ (Sv/Gy) |
|---|---|---|---|---|---|---|---|---|---|---|---|
| RQR 2 | 28.2 | 0.31 | 0.31 | 0.29 | 0.26 | 0.21 | 0.15 | 0.10 | 0.09 | 0.16 | 0.12 |
| RQR 3 | 32.4 | 0.41 | 0.40 | 0.38 | 0.34 | 0.28 | 0.20 | 0.14 | 0.14 | 0.22 | 0.17 |
| RQR 4 | 36 | 0.49 | 0.48 | 0.46 | 0.41 | 0.34 | 0.25 | 0.17 | 0.20 | 0.27 | 0.22 |
| RQR 5 | 39.4 | 0.56 | 0.55 | 0.52 | 0.47 | 0.39 | 0.29 | 0.20 | 0.25 | 0.32 | 0.25 |
| RQR 6 | 42.8 | 0.63 | 0.62 | 0.59 | 0.52 | 0.44 | 0.33 | 0.23 | 0.30 | 0.36 | 0.29 |
| RQR 7 | 46 | 0.69 | 0.68 | 0.65 | 0.58 | 0.49 | 0.37 | 0.26 | 0.35 | 0.41 | 0.33 |
| RQR 8 | 48.8 | 0.75 | 0.73 | 0.70 | 0.62 | 0.53 | 0.40 | 0.29 | 0.39 | 0.45 | 0.36 |
| RQR 9 | 53.9 | 0.84 | 0.82 | 0.78 | 0.70 | 0.60 | 0.46 | 0.33 | 0.46 | 0.51 | 0.41 |
| RQR 10 | 61 | 0.94 | 0.93 | 0.88 | 0.79 | 0.68 | 0.53 | 0.39 | 0.55 | 0.59 | 0.47 |
| RQA 2 | 31.5 | 0.44 | 0.44 | 0.41 | 0.37 | 0.30 | 0.22 | 0.14 | 0.15 | 0.23 | 0.18 |
| RQA 3 | 39.1 | 0.72 | 0.70 | 0.67 | 0.60 | 0.50 | 0.37 | 0.26 | 0.32 | 0.41 | 0.32 |
| RQA 4 | 46.1 | 0.95 | 0.93 | 0.89 | 0.79 | 0.67 | 0.51 | 0.36 | 0.50 | 0.57 | 0.45 |
| RQA 5 | 52.3 | 1.12 | 1.09 | 1.04 | 0.93 | 0.80 | 0.61 | 0.44 | 0.64 | 0.69 | 0.55 |
| RQA 6 | 58.2 | 1.24 | 1.21 | 1.16 | 1.04 | 0.89 | 0.69 | 0.51 | 0.75 | 0.78 | 0.62 |
| RQA 7 | 63.2 | 1.30 | 1.28 | 1.22 | 1.10 | 0.95 | 0.74 | 0.55 | 0.82 | 0.84 | 0.67 |
| RQA8 | 67.8 | 1.34 | 1.32 | 1.26 | 1.13 | 0.98 | 0.77 | 0.57 | 0.87 | 0.88 | 0.70 |
| RQA 9 | 76.5 | 1.37 | 1.35 | 1.28 | 1.16 | 1.01 | 0.81 | 0.60 | 0.91 | 0.91 | 0.73 |
| RQA 10 | 88.1 | 1.36 | 1.34 | 1.27 | 1.17 | 1.02 | 0.82 | 0.61 | 0.93 | 0.92 | 0.74 |
| RQT | 70.6 | 1.32 | 1.30 | 1.23 | 1.12 | 0.97 | 0.76 | 0.57 | 0.85 | 0.86 | 0.69 |

## 5. Discussion

Tables 1 - 5 contain important information for instrument designers, manufacturers and calibration services, The conversion coefficients listed here allow converting a measured or calculated value of the radiation field quantity kerma in air to the operational dose quantities $H^*$ and $H_p(10)$ proposed by ICRU RC 26. This permits to evaluate the suitability of existing radiation protection dosimeters and survey instruments for assessing the new quantities.

In comparison with the conversion coefficients to the present operational quantities $H^*(10)$ and $H_p(10)$ from [11], [12], the new conversion coefficients are lower, as could be expected from the monoenergetic conversion coefficients. Figure 7a shows the present and new conversion coefficients for the qualities in ISO 4037-1. The conversion coefficients for the x-ray qualities are plotted at the ordinate of their average fluence spectral energy $E_{avg}$. The data points are situated in the vicinity of the monoenergetic curves, which are given for comparison. Closer inspection shows that the conversion coefficients for the N- and L- series fall nearly exactly on the curves, which means that they are equal to the monoenergetic conversion coefficients at $E_{avg}$. Figure 7b is a similar representation of the conversion coefficients for diagnostic x-ray qualities from IEC 61267. For these broad spectra, substitution of the spectral conversion coefficient with the one for $E_{avg}$.is generally not possible.

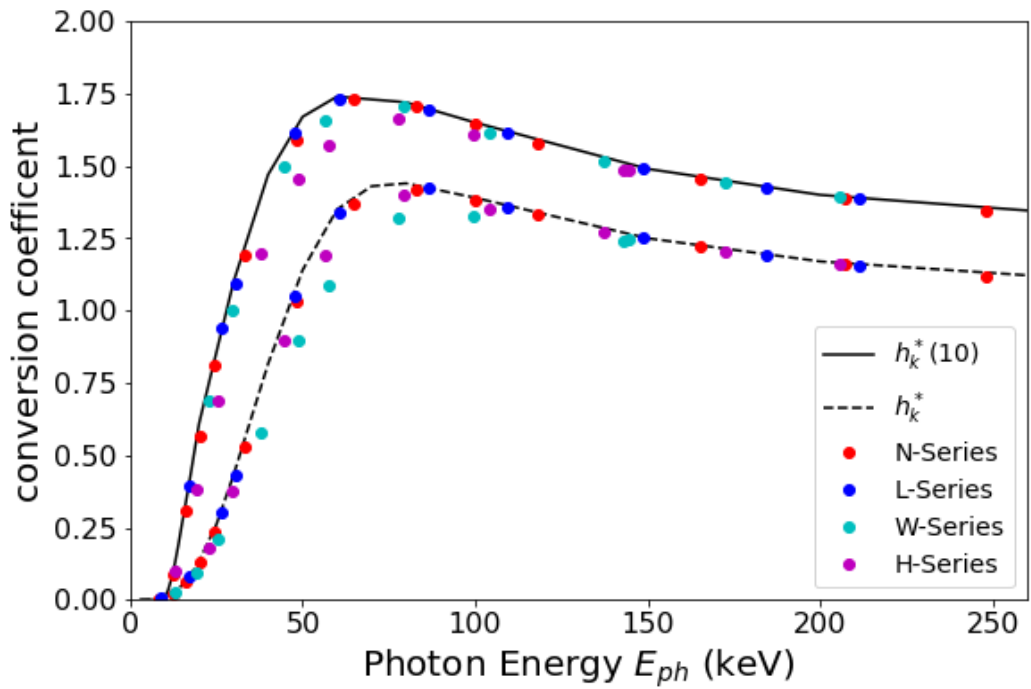

**Figure 7a:** Monoenergetic conversion coefficients for the present operational quantities $H^*(10) = H_p(10, 0^0)$ (full line) and for the new operational quantities $H^* = H_p(0^0)$ (dashed line). The data points grouped around the monoenergetic lines are the spectrum averaged conversion coefficients for the x-ray qualities from ISO 4037-1 [8] for the present and new operational quantities.

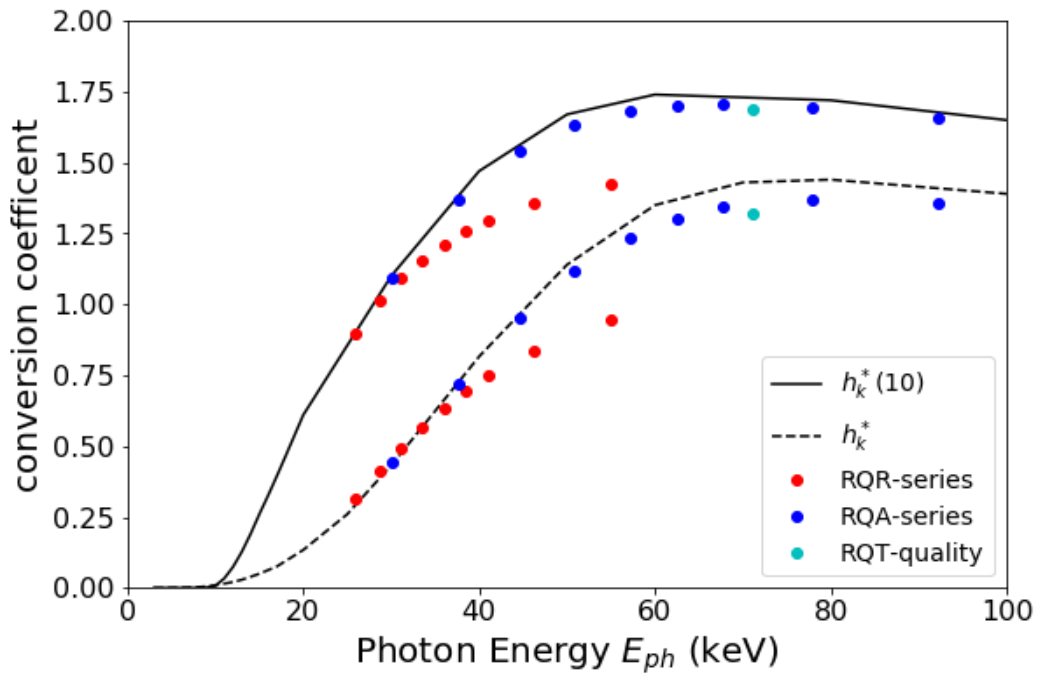

**Figure 7b:** Monoenergetic conversion coefficients for the present operational quantities $H^*(10) = H_p(10, 0^0)$ (full line) and for the new operational quantities $H^* = H_p(0^0)$ (dashed line). The data points grouped around the monoenergetic lines are the spectrum averaged conversion coefficients for the x-ray qualities from IEC 61267 [9] for the present and new operational quantities.

## Acknowledgments

Discussions with the members of ICRU Report Committee are gratefully acknowledged: N. E. Hertel, D. T. Bartlett, R. Behrens, J.-M. Bordy, A. Endo, G. Gualdrini and M. Pellicioni.
This research did not receive any specific grant from funding agencies in the public, commercial, or not-for-profit sectors.